\documentclass[letterpaper,12pt]{article}
\usepackage[margin=1in]{geometry}
\usepackage{setspace}
\usepackage{parskip}
\usepackage{courier}
\usepackage{caption}
\usepackage{subcaption}

\usepackage{amsmath}
\usepackage{amssymb}
\usepackage{graphicx}
\usepackage{algorithm}
\usepackage{amsthm}
\usepackage{multicol}
\DeclareGraphicsExtensions{.pdf,.png,.jpg}

\def\IF{{\bf if}\ }

\def\ELSE{{\bf else}}
\def\WHILE{{\bf while}\ }
\def\FOR{{\bf for}\ }

\def\blo{\noindent
\begin{tabular}{@{\quad}l@{\quad}}
\begin{minipage}{1in}
\begin{tabbing}
\qquad\=\qquad\=\qquad\=\qquad\=\qquad\=\qquad\=\qquad\=\kill}
\def\elo{\end{tabbing}\end{minipage}\\\end{tabular}}
\newenvironment{pseudocode}{\blo}{\elo}

\makeatletter
\newcommand*{\textoverline}[1]{$\overline{\hbox{#1}}\m@th$}
\makeatother

\newcommand{\Stash}{\mathcal{S}}

\begin{document}

\title{Memory Hierarchy Design for Caching Middleware in the Age of NVM
\thanks{Sandy Irani and Jenny Lam are with the University of California, Irvine.  Their research is supported in part by the NSF grant CCF-0916181.  A shorter version of this paper appeared in the IEEE 34th International Conference on Data Engineering (ICDE), Paris, France, 2018, pp. 1380-1383, doi: 10.1109/ICDE.2018.00155.}}

\author{
        {\em Shahram Ghandeharizadeh, Sandy Irani, Jenny Lam}
        \\
        \small Database Laboratory Technical Report 2015-01 \\
        \small Computer Science Department, USC \\
    \small Los Angeles, California 90089-0781 
}

\date{}

\maketitle


\begin{abstract}
Advances in  storage technology have introduced 
Non-Volatile Memory, NVM, as a new storage medium.
NVM, along with Dynamic Random Access
Memory (DRAM), Solid State Disk (SSD), and Disk present
a system designer with a wide array of options in designing caching middleware.
Moreover, design decisions to replicate a
data item in more than one level of a caching memory hierarchy
may enhance the overall system performance with a faster recovery time in the event of a memory failure.
Given a fixed budget, the key configuration questions are:
Which storage media should constitute the memory hierarchy?
What is the storage capacity of each hierarchy?
Should data be replicated or partitioned across the different levels of the hierarchy?
We model these cache configuration questions as an instance of the
Multiple Choice Knapsack Problem ({\sc MCKP}).
This model is guided by the specification of each type of memory
along with an application's database characteristics and its workload.
Although {\sc MCKP} is NP-complete, its linear programming relaxation is efficiently
solvable and can be used to closely approximate the optimal solution.
We use the resulting simple algorithm to evaluate design tradeoffs in 
the context of a memory hierarchy for a Key-Value Store (e.g., memcached)
as well as a host-side cache (e.g., Flashcache).
The results show selective replication is appropriate with certain failure rates
and workload characteristics.
With a slim failure rate and
frequent data updates, tiering of data across the different storage media
that constitute the cache is superior to replication.
\end{abstract}




\section{Introduction}\label{sec:intro}
The storage industry has advanced to introduce Non-Volatile Memory (NVM) such as
PCM, STT-RAM, and NAND Flash as new storage media.  
This new form of storage is anticipated to be much faster than Disk as permanent store and less expensive than DRAM as volatile memory.  When compared with DRAM, NVM retains its content in the presence of power failures and provides performance that is significantly faster than today's disk~\cite{ArulrajSigmod2017,Wu2016,ArulrajVLDB2016,Ou2016,Sun2015,ShiminVLDB15}.  While some NVM such as Memristor~\cite{memristor2008} are anticipated to be byte-addressable, others such as NAND Flash are block-based.


Caching middleware is an immediate beneficiary of NVM, challenging a system
designer with the following configuration questions:
\begin{enumerate}
\item Given multiple choices of storage media and a fixed monetary budget,
what combination of memory choices optimizes the performance of a workload?  
Is it appropriate to use one storage medium or a hierarchy of storage media?
 
\item With a hierarchy of storage media, should the system replicate or partition data across the storage medium?  
\end{enumerate}  
In this paper, we propose a framework that uses the characteristics of 
an application database and its workload to answer both questions.
Specifically, we devise an optimization problem that simultaneously optimizes 
both the configuration of the cache as well as
the decision as to whether to replicate data items on multiple storage media.
We demonstrate the use of this framework for two caching middleware:
an application-side cache and a host-side cache~\cite{cloudDM16}.
An application-side cache augments a data store with a key-value store (KVS) that enables an application to lookup the result of queries issued repeatedly.
It is designed for workloads that exhibit a high read to write ratio such as social networking.
An example KVS is memcached in use by Facebook that reports a 500 read to 1 write workload characteristic~\cite{bronson15}.

A host-side cache stages disk pages on a storage medium faster than
permanent store and brings data closer to the application.
With these caches, data is either a disk block, a file, or an extent
consisting of several blocks.
For example, Dell Fluid Cache~\cite{Ghandeharizadeh16} stages disk pages referenced by a database
management system such as MySQL on NAND Flash in order to hide the latency of
retrieving these pages from a significantly slower permanent store.
In~\cite{cloudDM16}, we report these two caches are in synergy with one another.
Moreover, we observe that the working set size to cache size ratio has a dramatic
impact on the choice of DRAM versus NAND Flash, motivating the aforementioned configuration 
questions.

In this paper, we use the term {\em cache} to refer to either one or several storage media used as a temporary staging area for data items.  A {\em data item} may be a key-value pair with the KVS or a disk page with the host-side cache.  Each type of storage medium that contributes space to the cache is termed a {\em stash}.  For example, in a KVS, the cache may
consist of DRAM and NVM as two stashes.  The stashes in a host-side cache may be PCM and Flash.  A copy of a data item occupying the cache is also stored on {\em permanent} store.  With memcached, the permanent store is a data store such as MySQL.  With 
a host-side cache such as Dell Fluid Cache, the permanent store may be a Storage Area Network (SAN) such as Dell Compellent.  
An application's read request for a data item fetches the data item from either the cache (a cache hit) or the permanent store (a cache miss) 
for processing.


We consider both {\em tiering} and {\em replication} of data items across stashes.
Tiering maintains only one copy of a data item across the stashes.
Replication constructs one or more copies of a data item across stashes.
Replication can be costly if the data item is updated frequently.
However, it expedites recovery of the data item in the presence of a stash failure, e.g., power failure with DRAM and hardware failure with NVM.  We capture this tradeoff in the formulated optimization problem and its solution.

The {\em primary contribution} of this paper is two folds.  
First, an offline optimal algorithm that computes (1) the choice and sizes of the
stashes that constitute a cache given a fixed budget and (2) 
a mechanism to determine when it is better to replicate an item across more than one
stash.
The input to the algorithm is the workload of an application (frequency of access to its referenced data items, sizes of the data items, and the time to fetch a data item on  a cache miss) and the characteristics of candidate stashes (read and write latency and transfer times,  failure rate, and price).  
The proposed algorithm uses the distribution of access frequencies to guide overall design choices in determining how much if
any of 
each type of storage media to use for the cache.  The algorithm can also be used
to guide high-level caching policy questions such as whether to maintain backup copies (replication) of data items in slower, more reliable storage media or whether to only keep a single copy of each item across stashes (tiering).

\begin{figure}
    \centering\small
        \includegraphics[width=8cm]{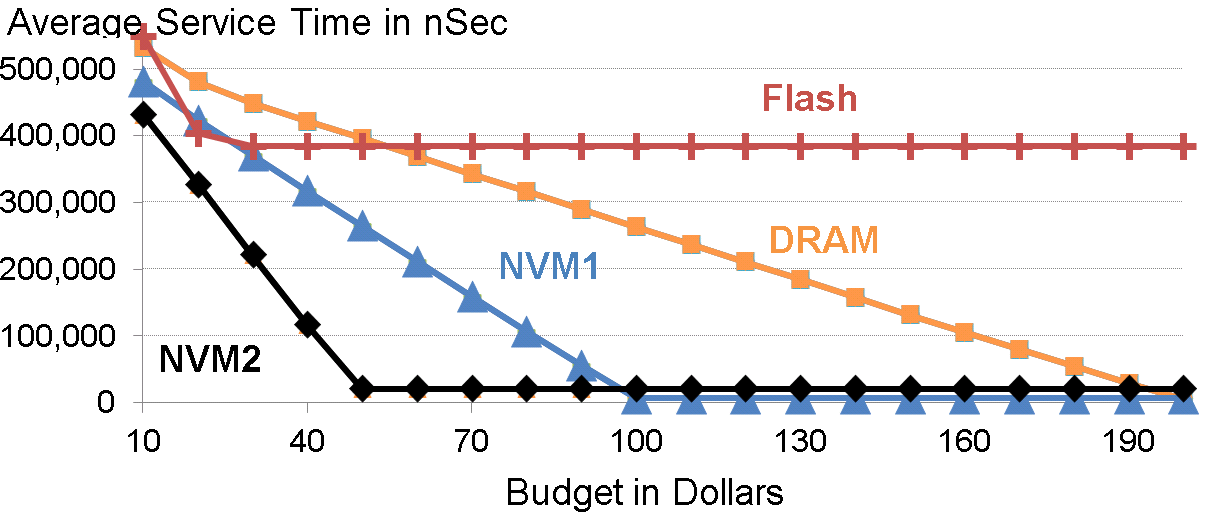}
\caption{Average service time of processing a social networking workload with different choice of storage medium for the cache.}
\label{fig:KVSTieringPerf1choice}
\end{figure}

For example, Figure~\ref{fig:KVSTieringPerf1choice} shows the estimated average service time as a function of budget when the cache is limited to a single stash. Each line corresponds to a  different storage medium used for the cache, with $NVM_1$
and $NVM_2$ corresponding to two representative NVM technologies (see Table \ref{tab:devices} for their characteristics).
With a tight budget (small x-axis values), NAND Flash is a better alternative than DRAM and comparable to $NVM_1$ because its inexpensive price facilitates a larger cache that enhances service time.  
Figure~\ref{fig:KVSTieringPerf1choice}  illustrates that except in the extreme case where the budget is large enough to store the entire database in DRAM, the two NVM options are preferable storage media to DRAM. This information would influence a design choice in determining which type of memory to use to cache key-value pairs. The algorithm is flexible and performs the same optimization to realize caches consisting of two or more stashes.  In addition, it evaluates choices in which data items are stored
in more than one stash.
Note that the algorithm uses the parameters of the storage technologies as input, so it can adapt to different types of storage media, depending on what is available.

Our method for cache configuration specifies a size for each stash in bytes. However, one typically purchases memory in certain granularities such as megabytes or gigabytes.  We assume that in determining the amount of memory for each stash, the byte figures would be rounded to the nearest value available for each memory type.

A second contribution is our trace driven evaluation of the proposed method.
The main lessons of this evaluation are:
\begin{enumerate}
\item 
Some combination of storage media perform considerably better than others over a wide range
of budget constraints. 
For example, the combination of Flash and $NVM_2$ is a good 
choice in the context of host-side caches for most budget scenarios
(See Figure \ref{fig:MailTierTwoTypes}). DRAM and $NVM_2$ does best for key-value stores
for most budgets.

\item
After a certain threshold point (that depends on database size and workload
characteristics),  spending money on  larger and faster caches
does not offer significant improvement in performance.

\item Before spending money on a pricey NVM that stores a small fraction of data items,
better performance gains might be achievable by purchasing a slightly slower speed 
storage medium with a higher storage capacity that can stage all data items.

\item Optional replication of data items by our algorithm produces results that are
slightly better than tiering and significantly superior to an approach that 
replicates all data items across stashes.

\item There can be many very different placements that approximate the optimal placement 
in their average service time.
These alternatives can be explored by limiting the set of placement options and comparing 
the average service time under the more restrictive scenario to the  
average service time in which all possibilities are allowed.  
\end{enumerate}

The rest of this paper is organized as follows.  
Section~\ref{sec:model} presents a model of the optimization
problem using the language of cache augmented data stores as it is more general.
Section~\ref{sec:mckp} formalizes this optimization problem as an instance of the
{\sc Multiple Choice Knapsack Problem} and 
presents a near optimal algorithm to solve it.
Section~\ref{sec:eval} demonstrates the effectiveness of this algorithm in
deciding the choice of stashes and placement of data items across them
for host-side caches using a trace-driven simulation study.
The results for cache augmented data stores are given in the appendix.
We describe related work in Section~\ref{sec:related} and detail brief future work in Section~\ref{sec:future}.

\section{The Model}
\label{sec:model}

We describe the model using the language of key-value stores because it is the most general. At the
end of Section \ref{sec:model}, we describe the few changes to adapt our methods for designing a  host-side cache.

We model query  sequences by a stream of independent events in which the occurrence of a particular query  
(key-value read request) or update (key-value write request) does not change the likelihood that a different query or update occurs in the near future. Independently generated events is the model employed by social networking benchmarks such as BG~\cite{sumita13} and LinkBench\cite{linkbench13}. If the probabilities of queries and updates to each key-value pair
are known a priori, then a static assignment of key-value pairs to each storage medium is optimal. Before each request, the optimal placement minimizes the expected time to satisfy the next request. If the distribution does not change over time, then the same placement will be optimal for every request. 
In practice, the placement of data items to stashes would be periodically adjusted to reflect changing access patterns.
In our simulation studies, we estimate the probability that a particular key-value pair is requested by analyzing the  frequency of requests to that
key-value pair in the trace file. We then determine an optimal memory  configuration based on those probabilities.

For each key value pair $k$, we use the following
four quantities\footnote{Note that the outcome of the algorithm (the quantities and types of storage to buy in designing a cache) does not depend critically on the read and write frequencies of individual items as these items can be moved to different stashes over time. Instead, the outcome of the algorithm depends on the total size of items with similar read/write characteristics. The idea is to estimate the characteristics of the load using the collective history for the individual items.}: 
\vspace{.1in}
\begin{itemize}
\item $size(k)$ the size of the key-value pair in bytes.
\item $comp(k)$ the time to compute the key-value pair from the database.
\item $f_R(k)$ the frequency of a read reference for key $k$.
\item $f_W(k)$ the frequency of a write reference for key $k$.
\end{itemize}
\vspace{.1in}

Note that $\sum_{k} (f_R(k)+f_W(k)) =1$.

There is a set of $\mathcal{S}$ candidate stashes for the cache,
each made of a different memory type.
For example, Table~\ref{tab:devices} shows
the memory device types\footnote{Since parameters for emerging NVM technology are not completely known,
we used two representative NVM types, which we call $NVM_1$ and $NVM_2$.} and their parameters used in our simulations.
Hence, in our experimental studies for the KVS, $\mathcal{S}$ is defined as:
$\mathcal{S} = \{
\mbox{Disk},
\mbox{Flash},
NVM_2,
NVM_1,
\mbox{DRAM}
\}$.
Each stash $s \in \Stash$, has the following characteristics:
\vspace{.1in}
\begin{itemize}
\item
$\delta_{R,s}$ the read latency of candidate stash $s$.
\item
$\delta_{W,s}$ the write latency of candidate stash $s$.
\item
$\beta_{R,s}$ the read bandwidth of candidate stash $s$.
\item
$\beta_{W,s}$ the write bandwidth of candidate stash $s$.
\item
$pricePerByte(s)$ the monetary cost of purchasing a byte of $s$.
\end{itemize}
\vspace{.1in}
The time to read $k$ from $s$ is
$T_R(s, k) =
\delta_{R,s} + size(k)/\beta_{R,s}$.
The time to write $k$ to $s$ is
$T_W(s, k) =
\delta_{W,s} + size(k)/\beta_{W,s}$.

\subsection{Placement Options}

A copy of a key-value pair $k$ can be placed in one or more of the stashes.
We define a {\em placement option} $P$ for a key-value pair to
be a subset of the set of stashes.
For example, the placement option $P = \{Flash, DRAM\}$
represents having a copy of a key value pair on both Flash and DRAM.
The placement option $\emptyset$ represents the scenario where a key-value
pair is not stored in the cache at all.
In each experiment, we define a set of possible placement options
for a key-value pairs that allows us to study the trade-offs between different
options. For example, in tiering, there is at most one copy of a key value pair
in the entire cache, so  the collection of possible placements would be
$\emptyset$, $\{\mbox{Disk}\}$, $\{\mbox{Flash}\}$, $\{NVM_2\}$, $\{NVM_1\}$, and $\{\mbox{DRAM}\}$.
In examining the trade-off between tiering and replication for a system with Flash and DRAM,
the set of possible placements would be
$\emptyset$, $\{\mbox{Flash}\}$, $\{\mbox{Disk}\}$, and $\{\mbox{Flash, Disk}\}$.

We define an optimization problem that minimizes the expected time per request
given a set of possible placement options and budget to purchase the memory for each stash.
For each key-value pair $k$ and each possible placement option $P$, there is an indicator
variable $x_{P, k} \in \{0, 1\}$, indicating whether $k$ is placed according to $P$.
If $P = \{Flash, DRAM\}$ and $x_{P, k} = 1$,  
then we are using replication with a copy of $k$ on both the Flash stash
and the DRAM stash. The  constraint $\sum_{P} x_{P, k} = 1$ says that $k$ has exactly one placement,
where the sum is taken over all possible placement options, including $\emptyset$.

If $x_{P, k} = 1$, then we must purchase $size(k)$ bytes of memory for each
stash in $P$. Again, if $P = \{Flash, DRAM\}$, we need $size(k)$ bytes of
Flash and $size(k)$ bytes of DRAM for the copies of key-value pair $k$.
Thus, the monetary price of having key-value pair $k$ in placement option $P$ is
$price(P,k) = \sum_{s: s \in P} size(k) \cdot costPerByte(s)$.
If the overall budget is $M$, then the total cost, summed over all key-value pairs
must be at most $M$:
$$\sum_k \sum_{P}  x_{P, k} \cdot price(P, k)  \le M.$$


\subsection{Expected Service Time}\label{sec:expserv}

The objective of the optimization is to expedite the average processing time of a request.  
This translates to minimizing the average service time.
For a key-value pair $k$ and placement option $P$, we define $serv(P,k)$
as the average service time of requests referencing $k$ if $k$ 
is assigned to stashes specified by the placement $P$.
The placement option $P = \emptyset$ is a special case because it does not assign $k$ to the cache, requiring every read reference to compute $k$ using the data store (i.e., the permanent store) and incur its service time $comp(k)$.
In this case, $serv(\emptyset, k) = f_R(k) \cdot comp(k)$.

For $P \neq \emptyset$,
there are three components to the service time for a key value pair $k$ if it is
placed according to $P$:   $(1)$
the time spent reading $k$, $(2)$ the time spent writing $k$, and $(3)$ the average cost of
restoring the copies of $k$ to a failed stash after repair. 
We consider
each in turn. 

If key-value pair $k$ is assigned according to $P$, then  upon a read request
to $k$, it is read from the stash with the fastest read time for $k$:
$\Delta_R (P, k) = \min_{s \in P} T_R (s, k)$.
The average service time to read $k$ is its read 
frequency 
times the time for the read: $f_R(k) \Delta_R (P, k)$.

Upon a write to $k$, all the copies of $k$ across the stashes
dictated by $P$ must be updated\footnote{An alternative definition is
to assume concurrent writes to all copies with the slowest stash
dictating the time to write $k$:
$\Delta_W (P, k) = \max_{s \in P} T_R (s, k)$.  A strength of the proposed
model is its flexibility to include alternative definitions.}:
$\Delta_W (P, k) = \sum_{s \in P} T_W (s, k)$.
The average time writing $k$ will be the frequency of a write request to $k$
times the time for the writes: $f_W(k) \Delta_W (P, k)$.

We model failures as a rate $\lambda$ that defines the inter-arrival
between two failures in terms of the number of requests as $\frac{1}{\lambda}$.
For example, a failure rate of 0.001 ($\lambda$=0.001) means that the average number of
requests between occurrences of two failures is 1000.  
We define a failure event $F$
as the set of stashes that fail at the same time.
To simplify discussion, we assume $F$ consists of one failure (i.e., only one stash fails at a time).
However, the model is general enough to express the optimization problem
in its full generality that incorporates every possible subset $F$ of stashes failing.
For each such failure event, we determine the cost of restoring the contents of the impacted
stash. 
We require a failure rate $\lambda_F$ for every possible failure event $F$. 
Section~\ref{sec:failRate} details how we compute $\lambda_F$ in our experiments
using both trace files and parameter settings of stashes.

In the presence of failures, it may be advantageous to store a key-value pair $k$ on more than one
stash. If $k$ is stored on two stashes and one  fails, then the time to repopulate
the failed stash is reduced by retrieving a copy of $k$ from the other stash.
This is the only 
motivation for replicating a key-value pair across more than one stash. 
Having an extra copy of a key-value
pair increases the cost of updating on writes. 
This trade-off between the cost
of updating an additional copy and the benefit of having the extra copy in case of failure 
determines how many copies of a key-value pair is optimal. 
The optimization problem we define here automatically takes these considerations into account.

For each triple $(F, P, k)$, we define 
$fail$, the cost of restoring $k$ after a 
failure event $F$ given that $k$ is stored according to placement option $P$.
$failCost(F, P, k) = 0$ if the set of stashes that fail 
has no overlap with $P$ (i.e., $P \cap F = \emptyset$). 
Otherwise, there are two components to the cost: 
First, the cost of retrieving
a copy of $k$. 
If all stashes of $P$ are wiped clean after a failure event $F$ (i.e. $P \subseteq F$),
then
$retrievalCost(F, P, k) = comp(k)$. Otherwise
$k$ is read from the fastest stash still available:
$retrievalCost(F, P, k) = \min_{s \in P-F} \Delta_R(s, k)$.

The second component of the incurred cost after a failure
is restoring $k$ to the stashes in $P$ that 
failed during failure event $F$:
$restoreCost(F, P, k) = \sum_{s \in P \cap F} \Delta_W(s, k)$.

Now, we assemble all components of the service
time of a request referencing a key-value pair $k$ assigned
according to the placement option $P$:

\begin{eqnarray*}
serv(P,k) & = & f_R(k) \Delta_R (P, k)
 +  f_W(k) \Delta_W (P, k)\\
 & & \\
& + & \sum_{F} \lambda_F  \cdot restoreCost(F, P, k) \\
& + & \sum_{F} \lambda_F \cdot retrievalCost(F, P, k)
\end{eqnarray*}

The goal is to select values for the variables $x_{P,k} \in \{0, 1\}$
that minimizes
$$\sum_P \sum_k x_{P, k} \cdot serv(P, k)$$
In the next section, we show how this optimization problem
can be solved using known techniques for the Multiple Choice
Knapsack Problem. Knapsack problems are typically maximization
problems, so we define an equivalent maximization problem
to the problem stated above which maximizes a benefit
instead of minimizing a cost.
For each placement $P$ and key-value pair $k$, 
define
$ben(P, k) = serv(\emptyset, k) - serv(P, k)$. 
Note that the benefit  of placement
$\emptyset$ is $0$ (i.e., $ben(\emptyset,k)=0$). The constraints and definitions for the problem are unchanged, except
that the goal is now to select a placement for each $k$ that maximizes
$\sum_P \sum_k x_{P, k} \cdot ben(P, k)$.
An optimal solution to the minimization problem is also an optimal
solution for the maximization problem and vice versa.

To summarize,  the 
{\sc Cache Configuration Problem} 
 is to find a placement
for each $k$
(i.e., values for the indicator variables $x_{P,k} \in \{0, 1\}$)
that maximizes
$\sum_P \sum_k x_{P, k} \cdot ben(P, k)$,
subject to the constraints that for each $k$,
$\sum_{P} x_{P, k} = 1$,
and for budget $M$,
$$\sum_k \sum_{P}  x_{P, k} \cdot price(P, k)  \le M$$
Every sum over $P$ is taken over the set of possible placement
options that we define for a particular experiment (i.e., instance
of the optimization problem). 
The values of the indicator variables dictate the size of 
the stashes: each stash must be large enough to hold a copy
of every key-value pair that has a copy on that stash.
For a stash $s$, we must sum over all placement options
that include a copy of a key-value pair on $s$. The total
size of stash $s$ is then
$\sum_{P: s \in P} \sum_k x_{P, k} \cdot size(k)$.

\subsection{Host-Side Caches}

We use the same framework to optimize a configuration for host-side caches.
There are two key conceptual differences that are accommodated using our logical formalism.
First, today's host-side caches uses Flash (instead of DRAM of cache
augmented data stores, e.g., memcached)
and it is natural to extend them with NVM.
Hence, the set of stashes to consider are
$\mathcal{S} = \{
\mbox{Flash},
NVM_1,
NVM_2
\}$.
Second, a data item $k$ might be either a disk page or a file.
The cost of not assigning $k$ to the cache means it must be serviced using the permanent store (might be a Disk controller, a RAID, a Storage Area Network):   
$$serv(\emptyset, k) = f_R(k) \cdot T_R(Disk, k) + f_W(k)
\cdot T_W(Disk, k).$$
$T_R(Disk, k)$ and $T_W(Disk, k)$ are the time to
read and write $k$ to Disk.
Other definitions of Section \ref{sec:model} are unchanged.

\section{The Multiple Choice Knapsack Problem}\label{sec:mckp}

In the {\sc Knapsack Problem},
there is a set  of items available to pack into a knapsack.
Each item has a benefit and a weight.
The knapsack has a weight limit and
the goal is to select the set of items to pack into the knapsack
that  maximizes 
the total benefit of the items selected and does not exceed
the weight limit of the knapsack.
In the {\sc Multiple Choice Knapsack Problem},
the items are partitioned into groups with the additional constraint that
at most one item from each group can be selected. 

In the {\sc Cache Configuration Problem}, 
each key-value pair has a set of placement options 
(including the option of not placing it on any of the 
memory banks, reading it from permanent store always). Therefore each key-value pair defines a class
from which we are selecting at most one option. 
Each selection has an associated benefit (as defined
in Section \ref{sec:model}) and each selection has a price
that corresponds to the ``weight" of the choice in the knapsack
problem. Therefore the {\sc Cache Configuration Problem}
 can be cast as an instance of {\sc MCKP}.

The Knapsack Problem is a well-studied problem in the
theory literature and is  known to be NP-hard \cite{GareyJohnson}.
Since the Knapsack problem is a special 
case of  {\sc MCKP} in which each item is in a category of its own,
{\sc MCKP} is also NP-hard.
The books \cite{Martello:1990:KPA:98124} and \cite{KelPfePis04} are dedicated to the Knapsack Problem and its
variants and both include a chapter on {\sc MCKP}.
One can consider a linear programming relaxation in which the 
indicator variables $x_{p, S}$ can be assigned real values in the range from
$0$ to $1$ instead of $\{0, 1\}$ values. 
In the case of {\sc MCKP}, the LP relaxation can be optimally
solved by the following greedy algorithm:
start with the placement $\emptyset$ for all of the key-value pairs.
This placement has $0$ overall benefit and $0$ monetary cost.
The algorithm considers a sequence of changes  (to be defined later)
in which the placement of a key-value pair is upgraded to a
more beneficial and more costly placement option.
This process continues until the money runs out
and the last item can only be partially upgraded.
The solution obtained by the greedy algorithm is optimal
for the LP-relaxation  of the problem \cite{SZ} and therefore at least
as good as the optimal integral solution. Moreover, only one
item is fractionally placed. The algorithm used here follows the
greedy algorithm but stops short of the last upgrade that results
in a fractional placement. Let $OPT_{frac}$ and $OPT_{int}$
represent the total benefit obtained by the
optimal solutions for the fractional and integer versions
of {\sc MCKP} respectively. Let
$GR_{frac}$ and $GR_{int}$
represent the  benefit obtained by the greedy algorithm
for the fractional and integer versions
of {\sc MCKP} respectively. We have:
$$GR_{int} \le OPT_{int} \le OPT_{frac} = GR_{frac}.$$
Moreover, the difference in overall benefit between
the greedy integral  solution and the greedy fractional solution is at most
the benefit obtained by
the last fractional upgrade. Since individual key-value pairs are small
with respect to the overall database size, the effect of not
including the last partial upgrade is not significant.

The running time of the algorithm depends on 
  $n$,  the number of key-value pairs, and  $p$,  the
number of different placement options considered.
With replication, 
the largest $p$ is $2^m$ where $m$ is the number of different
stashes since any subsets of the stashes could be a placement option.
For tiering, $p = m+1$ because each placement option is a single 
stash. The additional $1$ comes from the $\emptyset$ option.
Our implementation uses a priority queue to select the
next upgrade resulting in an overall running time of
$O(np \log np)$.
There are more complex algorithms to find the greedy solution that run 
in time $O(np)$ \cite{Dyer84, Zemel:1984:OAL:441.491}. 
We chose  the $O(np \log np)$ implementation because it
was easier to implement.

We now give a description of the greedy algorithm
for {\sc MCKP}.
The $p$ different placement options,
 $P_0, \ldots, P_{p-1}$,
  are sorted in increasing
 order by price, so $price(P_i,k) \le price(P_{i+1},k)$.
 The option $P_0$ is the $\emptyset$ option, and
  $price(P_0,k) = ben(P_0,k) = 0$.

Not every placement option is a reasonable choice for every key
value pair. For example, if placement option $P$ for key $k$
has lower benefit and higher price than option $P'$ for
$k$, then $P$ should not even be considered as a viable option
for key $k$.
  Thus, the  first step is a preprocessing step
in which  a list of viable options is determined for
  each $k$. If $j$ is not on $k$'s viable list, then
  $P_j$ will never be considered as an option for $k$.
The pseudo-code for  {\sc SetViableOptions} is given 
in Algorithm \ref{alg:SVO}.

\begin{algorithm}
\begin{small}
\begin{pseudocode}
Initialize {\tt viableList(k)} to be an empty list\\
Initialize {\tt done} = false\\
Initialize {\tt curr} = 0\\
\WHILE (not {\tt done})\\
\> {\tt viableList(k).add(curr)}\\
\> {\tt maxGradient} = $-\infty$\\
\> \FOR  $j = $ {\tt curr}+1 $\ldots p-1$\\
\>\> {\tt deltaBen} = $ben(P_{curr},k) - ben(P_j,k)$\\
\>\> {\tt deltaPrice} = $price(P_{curr},k) - price(P_j,k)$\\
\>\> g = {\tt deltaBen}/{\tt deltaPrice}\\
\>\> \IF (g $>$ {\tt maxGradient}) \\
\>\>\> {\tt maxGradient} = g \\
\>\>\> {\tt next} = j \\
\> \IF ({\tt maxGradient} $>$ 0)\\
\>\>{\tt curr} = {\tt next}\\
\>\ELSE \\
\>\> {\tt done} = true
\end{pseudocode}
\end{small}
\caption{SetViableOptions(k).}
\label{alg:SVO}
\end{algorithm}

\begin{figure}[t]
\begin{center}
\includegraphics[width=6.5in]{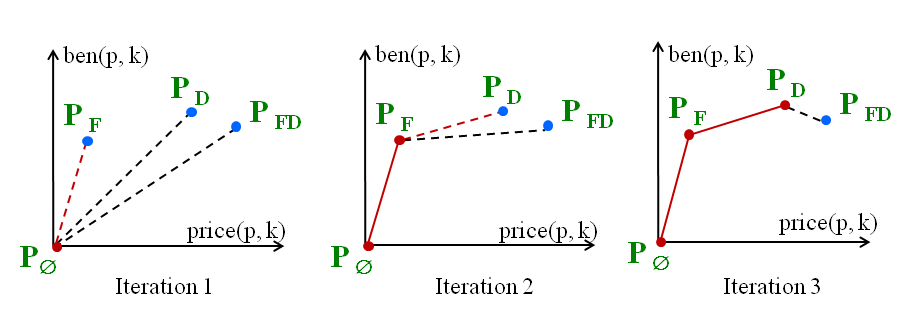}
\caption{Illustration of {\sc SetViableOptions}. Before the first iteration, the viableList for $k$ is initialized to  $[ \emptyset ]$.
        The algorithm examines each segment connecting $P_{\emptyset}$
        to the three placements to the right and selects $P_F$ because the segment
        from $P_{\emptyset}$ to $P_F$ has
        the largest slope. The viableList for $k$ is now  $[ \emptyset, \mbox{F} ]$.
        In the second iteration, the algorithm looks at the segments connecting $P_{F}$
        to the two placements to the right and selects $P_D$ because the segment connecting $P_{F}$ to $P_D$ has the largest slope.
The viableList for $k$ is now $[ \emptyset, \mbox{F}, \mbox{D}]$.
        FD is not added to $k$'s viable list in the third iteration because the slope
        from $P_D$ to $P_{FD}$ is negative, indicating that placing $k$ on Flash and DRAM
        costs more money and brings less benefit than placing $k$ on DRAM only.}
\label{fig:Iterations}
\end{center}
\end{figure}

The procedure {\sc SetViableOptions} is illustrated graphically
in Figure \ref{fig:Iterations}. For this example, only two
memory banks are considered: DRAM and FLASH. 
We consider four placement options for a key-value pair
$k$. The key-value pair can be placed on both DRAM and FLASH (FD),
Flash only (F), DRAM only (D), or neither ($\emptyset$).
Each placement option is represented as a point with
the horizontal axis representing the price of 
placing $k$ on that placement option and the vertical axis representing the
benefit. Essentially, a placement option is viable for $k$ if its corresponding
point lies on the convex hull of all the points and the slope of the
segment connecting it to the previous point is positive.

\begin{figure}[t]
\begin{center}
\includegraphics[width=5in]{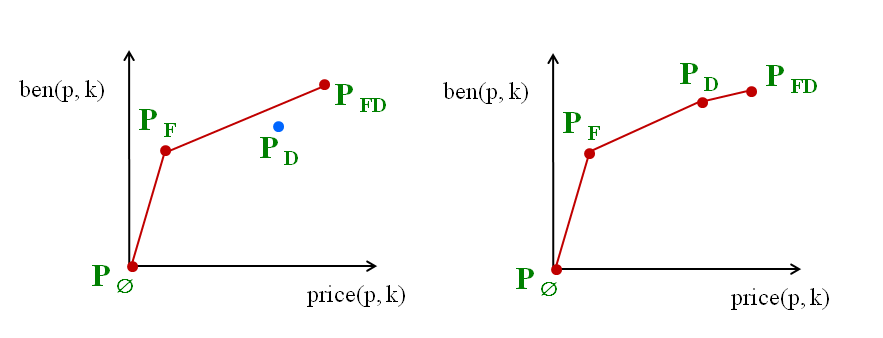}
\caption{The viableList for the key on the left is $\langle \emptyset, F, FD \rangle$.
       After $F$ was added, the segment from $P_F$ to $P_{FD}$
       had a larger slope than the segment from $P_F$ to $P_D$, so option $D$
was bypassed and $FD$ was added to $k$'s viable list. The viableList for the key
on the right  is $\langle \emptyset, F, D, FD \rangle$.}
\label{fig:OtherKeys}
\end{center}
\end{figure}

The fact that the point corresponding to $P_D$ is placed
higher than $P_{FD}$ for the particular key-value pair
$k$ used in Figure \ref{fig:Iterations}
means that $ben(P_D, k) > ben(P_{FD}, k)$. This arrangement may not
be the same for all key-value pairs $k$ and will in general, depend on
a number of different parameters, especially the write frequency of
$k$. The more frequently $k$ is written, the more costly it is to maintain
the extra copy of $k$ on Flash. 
Figure \ref{fig:OtherKeys} shows two other possible scenarios for
the list of viable placements for a key-value pair.

\begin{figure}[t]
\begin{center}
\includegraphics[width=7in]{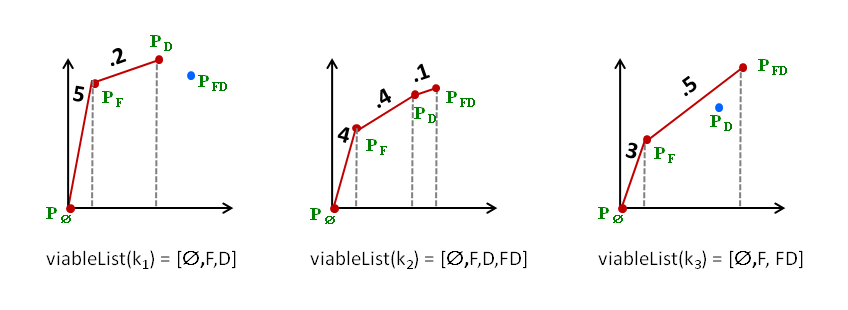}
\caption{Three key-value pairs and 
their placement options.
}\label{fig:greedyExample}
\end{center}
\end{figure}

After the viable list  for each key-value pair has been determined, the greedy algorithm initializes the placement
for each key-value pair to $\emptyset$.
In each iteration the greedy algorithm selects the key-value pair $k$ such that the slope of the segment from $k$'s
current placement to $k$'s next viable placement is maximized. Then greedy  upgrades the placement for $k$ to the
next placement option on its list of viable placements.
The process continues until the money runs out or until there is no upgrade that improves the overall
benefit (i.e., until all the upgrade gradients are negative). 
The pseudo-code for the greedy algorithm is given in Algorithm \ref{alg:greedy} which makes use of
a function that returns the upgrade gradient for a key value pair,
given in Algorithm \ref{alg:gug}.
The greedy algorithm is illustrated with a small example in Figure 
\ref{fig:greedyExample}.

\begin{table*}[t]
\begin{center}
\begin{small}
\begin{tabular}{|c||c|c|c|l|c|}
\hline
 & Segment  &     Key Value & Change in & Upgrade & \\
Order &  Slope &   Pair & Placement  & Description & Price\\
\hline
\hline
1 & 5 & $k_1$ & $\emptyset \rightarrow$ F & Assign $k_1$ to Flash & 1 \\
\hline
2 & 4 & $k_2$ & $\emptyset \rightarrow$ F & Assign $k_2$ to Flash & 1\\
\hline
3 & 3 & $k_3$ & $\emptyset \rightarrow$ F & Assign $k_3$ to Flash & 1\\
\hline
4 & .5 & $k_3$ & F $\rightarrow$ FD & Replicate $k_3$ to DRAM & 4\\
\hline
5 & .4 & $k_2$ & F $\rightarrow$ D & Move $k_2$ from Flash to DRAM  & 3\\
\hline
6 & .2 & $k_1$ & F $\rightarrow$ D & Move $k_1$ from Flash to DRAM  & 3\\
\hline
7 & .1 & $k_2$ & D $\rightarrow$ FD & Replicate $k_2$ to Flash & 1\\
\hline
\end{tabular}
\end{small}
\caption{Different iterations of the GreedyPlacement algorithm with the three keys.
The order of upgrades is according to the slope of line segments shown in 
Figure~\ref{fig:greedyExample}.}
\label{tab:greedyUpgrades}
\end{center}
\end{table*}

\noindent {\bf Example:}  We illustrate the GreedyPlacement algorithm with 3 key-value pairs 
using two candidate stashes DRAM and Flash.
The graph for each key-value pair is shown in Figure~\ref{fig:greedyExample}.
Each point in the graph is a placement option with the horizontal axis representing
the price and the vertical axis representing benefit.
The graph for $k_1$ and $k_3$ consists of two segments while the one for $k_2$ 
consists of three segments.
The slope of a segment is shown by a value next to it.  
For example, the two segments of $k_1$ have a slope of 5 and 0.2, respectively.

The horizontal distance between the endpoints of the segments (distance between
the dotted vertical lines) is the price of the upgrade.
For example, with $k_2$, the price of an upgrade from FLASH to DRAM (distance
between the vertical lines $P_F$
and $P_D$) is higher than the upgrade price from DRAM to both FLASH and DRAM (distance
between the vertical lines $P_D$ and $P_{FD}$).

Table~\ref{tab:greedyUpgrades} shows the different iterations of the GreedyPlacement 
algorithm with a price of 1 for Flash and 4 for DRAM.  
We assume the size of the three key-value pairs is identical.
Table~\ref{tab:greedyUpgrades} shows how GreedyPlacement assigns according to segments with the highest
slope first, see column 2 labeled ``Segment Slope".
The highest slope segment belongs to $k_1$, transitioning its assignment from
$\emptyset$ to the placement option Flash.
This is repeated with the second and third highest slope segments, transitioning
the assignment of both $k_2$ and $k_3$ to Flash.   
The next highest slope segment is 0.5 (see row 4) and belongs to $k_3$. 
It corresponds to changing the placement from $k_3$ from
DRAM to DRAM and Flash which entails 
 replicating $k_3$ onto DRAM with an additional cost of 4,
see last column of Table~\ref{tab:greedyUpgrades}.
The next line segment (slope of 0.4 belonging to $k_2$) changes the placement of
$k_2$ from Flash to DRAM, resulting in a price of 3, i.e., cost of 4 for DRAM and saving of 1
by removing $k_2$ from Flash.
This process continues until the available budget is exhausted.

A budget of 11 enables the first five iterations of GreedyPlacement.
Its final placement will have $k_1$ on Flash,
$k_2$ on Flash and DRAM, and $k_3$ on DRAM.
A budget of 13 accommodates the sixth iteration to upgrade $k_1$ to DRAM.

In sum, a budget of 11 and 13 configures a cache with two stashes consisting of Flash and DRAM.  The size of each stash is dictated by the total size of its assigned objects.
{\hspace*{\fill}\rule{6pt}{6pt}\bigskip}

\begin{algorithm}
\begin{small}
\begin{pseudocode}
Initialize {\tt moneySpent} = 0\\
Initialize {\tt done} = false\\
\FOR all $k$\\
\> {\tt viableList(k).reset()}\\
\WHILE (not {\tt done})\\
\> Select $k$ with the largest upgrade gradient\\
\> \IF GetUpgradeGradient(k) $>$ 0\\
\>\> {\tt next} = {\tt viableList(k).getNext()}\\
\>\> {\tt current} = {\tt viableList(k).getCurrent()}\\
\>\> {\tt deltaPrice} = $price(P_{next}, k) - price(P_{curr}, k)$\\
\>\> \IF {\tt deltaPrice} + {\tt moneySpent} $\le  M$\\
\>\>\> {\tt moneySpent} = {\tt moneySpent} + {\tt deltaPrice}\\
\>\>\> {\tt viableList(k).advance()}\\
\>\> \ELSE \\
\>\>\> {\tt done} = true\\
\>\ELSE \\
\>\> {\tt done} = true
\end{pseudocode}
\end{small}
\caption{GreedyPlacement($M$ as total budget)}
\label{alg:greedy}
\end{algorithm}

\begin{algorithm}
\begin{small}
\begin{pseudocode}
\IF viableList(k).isAtEnd()\\
\>return $-\infty$\\
next =  viableList(k).getNext()\\
curr =  viableList(k).getCurrent()\\
deltaBen = $ben(P_{next}, k) - ben(P_{curr}, k)$\\
deltaPrice = $price(P_{next}, k) - price(P_{curr}, k)$\\
return deltaBen/deltaPrice
\end{pseudocode}
\end{small}
\caption{GetUpgradeGradient(k)}
\label{alg:gug}
\end{algorithm}

\section{Evaluation}\label{sec:eval}

In this section we used the algorithm presented in Section~\ref{sec:mckp} as
a tool to evaluate different mixes of stashes under varying budget constraints in
the context of KVS and host-side caches.
Table \ref{tab:devices}
shows the parameters for the five types of memory used in this study,
including their read and write latency, read and write bandwidth,
price in dollars per gigabyte and mean time between failures.
The actual parameters of current NVM technology are still undetermined, so we selected two
representative  NVM types to use in the study. 
The methods of Section~\ref{sec:mckp} can take
as input any set of storage devices and corresponding parameters. 

A summary of the lessons learned from this evaluation are presented in Section~\ref{sec:intro}.
The rest of this section is organized as follows.
In Subsection \ref{sec:failRate}, we show how to compute the failure rates using the trace file in combination 
with the parameter settings of a candidate storage medium (see Table~\ref{tab:devices}).  
Next, in Subsections \ref{sec:hostSide} and
\ref{sec:KVS} we present a trace-driven evaluation of host-side and KVS
caches in turn.

\subsection{Failure Rates}
\label{sec:failRate}

Recall from Section~\ref{sec:expserv} that the inter-arrival between two failures is quantified in terms of the number of requests as 
$\frac{1}{\lambda}$ where $\lambda$ is the rate of failures.
For example, $\lambda$=0.001 means that on the average, there are 1000 requests between two failure occurrences.   
This models two kinds of failures:  
1) power failures that cause a volatile stash such
as DRAM to lose its content, and 2) hardware failures that require a stash such as 
NVM to be replaced with a new one.  
The former is characterized by Mean Time Between Failure (MTBF) and the later is 
quantified using Mean Time To Failure (MTTF).  
Both model constant failure rates, meaning, in every time unit a failure has the 
same chance as any other time instance.
MTBF is used for power failure because it pertains to a condition that is repairable.
MTTF is used for devices because we assume they are non-repairable and must be 
replaced with a new one.  
Both incur a Mean Time To Repair (MTTR).
With power failure, MTTR is the time to restore the power and for the system to 
warm-up the volatile memory with data.  
With NVM device failure, MTTR is the time for the system operator to shutdown the server,
replace the failed NVM with a new one, restore the system to an operational state,
and incur the overhead to populate the new empty NVM with data.

We require a failure rate $\lambda_F$ for every possible failure event $F$. 
In our experiments, we only consider failure events
that consist of a single device failure based on the assumption that failure events are sufficiently
infrequent that it is unlikely that one memory device fails while another is still down.
In these studies, to determine $\lambda_F$ where $F$ consists of a single stash (i.e., $F = \{s\}$), we
multiply the number of requests per hour in the trace times 
$(MTTF + MTTR)$ for non-volatile memories and $(MTBF + MTTR)$ for DRAM. 
The number of requests per hour  is determined by dividing the total
number of requests in the trace by the time (in hours) over which the trace was gathered.
A more elaborate way of modeling failure rates would be required for failure events in which
more than one device fails.

\begin{table*}
\begin{center}
\begin{small}
\begin{tabular}{|l|c|c|c|c|c|} 
\hline
\hline
                    & $NVM_1$  & $NVM_2$ & DRAM  &  Flash  & Disk \\
\hline
\hline
Read Latency in ns ($\delta_R$)        & $30$  & $70$  & $10$ & $25000$ & $2 \times 10^6$ \\
Write Latency in ns ($\delta_W$)      & $95$ & $500$   & $10$ & $2 \times 10^5$ & $2 \times 10^6$ \\
Read Bandwidth in MB/sec  ($\beta_R$)            &  $10 \times 1024$ & $7 \times 1024$ & $10 \times 1024$ & $200$ &  $10$\\ 
Write Bandwidth in MB/sec  ($\beta_W$)   &  $5 \times 1024$ & $1 \times 1024$  &  $10 \times 1024$ & $100$ & $10$ \\
Price in dollars per Gig       & $4$  & $2$ & $8$ &$1$ & $.1$ \\
MTTF/MTBF  in hours     & $21875$ & $43776$ & $8750$ & $87576$ & $87576$ \\
MTTR     in hours   & $24$ & $24$ & $10$ & $24$ & $24$ \\
MTTF/MTBF + MTTR     in years   & $2.5$ & $5$ & $1$ & $10$ & $10$ \\
\hline
\hline
\end{tabular}
\end{small}
\caption{Parameter settings of storage medium used in experimental evaluation.}
\label{tab:devices}
\end{center}
\end{table*}

\subsection{Host-Side Cache for Mail Server}
\label{sec:hostSide}

Today's host-side caches employ one storage medium, namely Flash. 
This section considers an extension consisting of Flash, $NVM_1$ and $NVM_2$ as possible stashes.
For our analysis, we use 
the disk block trace from a production 
mail server used on a daily basis for one week~\cite{koller10} and several different production servers at Microsoft~\cite{kava2008}.   
It consists of 438 million requests to 14.7 million blocks. 
The total size of the requested blocks is 56.25 gigabytes. 
The cost of storing all the requested disk pages on the most expensive stash, $NVM_1$, 
is just under \$225. 

\begin{figure}
    \centering\small
    \begin{subfigure}{0.48\textwidth}
        \includegraphics[width=8cm]{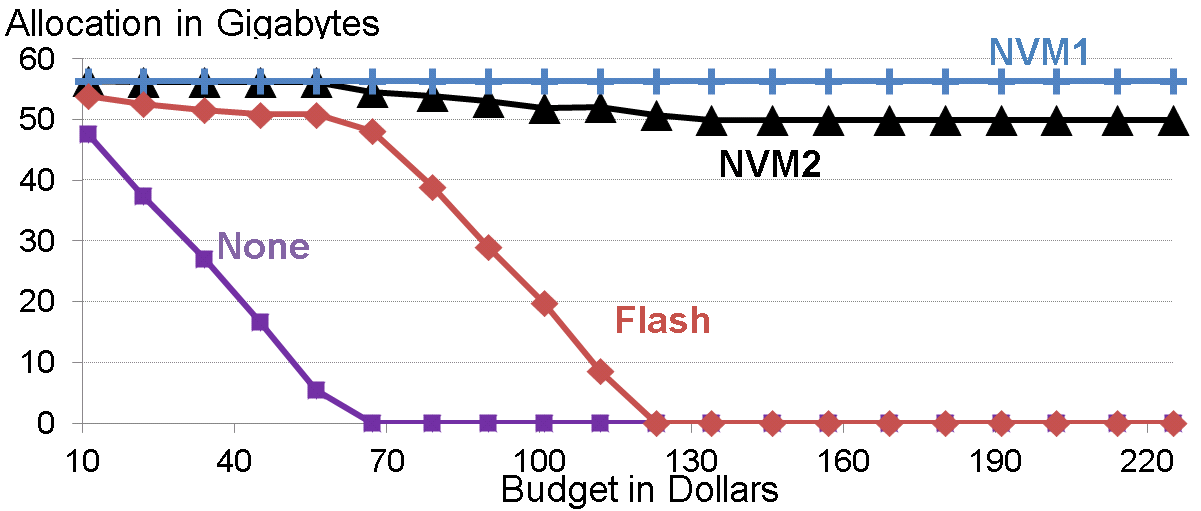}
        \caption{The cost of failures is included.}
        \label{fig:MailTierAllocFail}
    \end{subfigure}\vspace{2em}
    \begin{subfigure}{0.48\textwidth}
        \includegraphics[width=8cm]{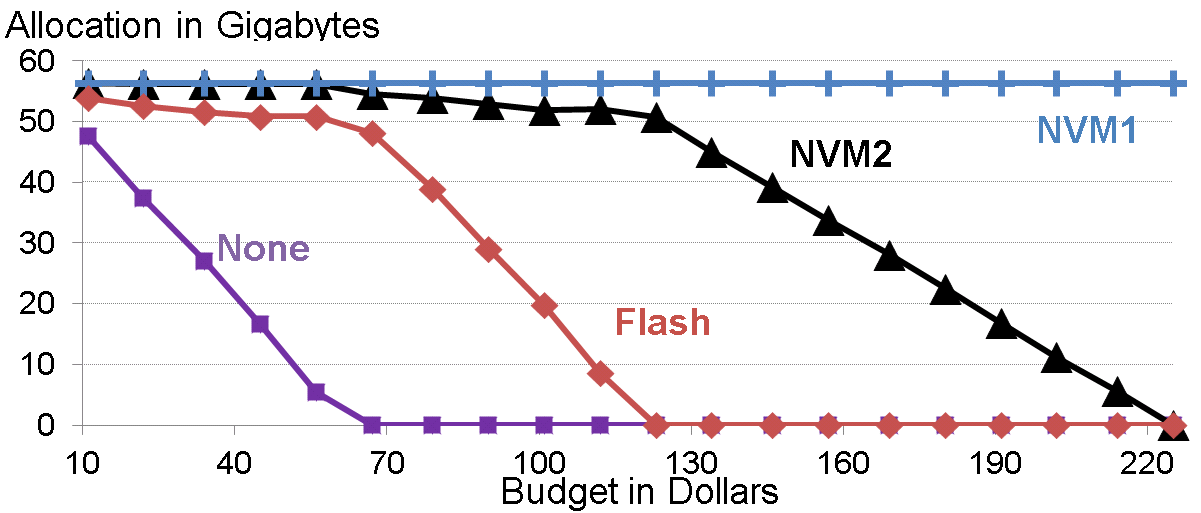}
        \caption{The cost of failures is not included.
        }
        \label{fig:MailTierAllocNoFail}
    \end{subfigure}\vspace{2em} 
   
        \caption{The optimal partition of the disk pages among the stashes as the budget varies.  The amount allocated 
to each stash is the vertical distance between the line labeled with  that memory type and the one below. 
In the first graph, the cost of failures is included. For budgets beyond \$124, most of the disk pages are in $NVM_2$.
In the second graph, the cost of failures is not included.
At the highest budget (\$225), all the disk pages are in $NVM_1$.
        }
\label{fig:MailTierAlloc}
\end{figure}

The first set of experiments examine different cache designs with a tiering policy in 
which each disk page is stored on at most one stash. In the first of these experiments, all three forms of memory 
(Flash, $NVM_1$ and $NVM_2$) are available as placement options. 
Figure \ref{fig:MailTierAllocFail} shows the allocation of blocks to stashes as the budget is increased up to \$225. 
For a particular budget, the size of each stash is the vertical distance between the line labeled with that stash and the line below. There is no budget at which the optimal allocation has  disk pages in $NVM_1$ and  disk pages not stored in the cache at all. In other words, before spending any money on upgrading disk blocks to $NVM_1$, it is more cost effective to get all of the disk blocks into the cache. There is sufficient variation in request frequency, however, for budgets in the range of \$100, it is optimal to have disk pages spanning Flash, $NVM_2$, and $NVM_1$. 

Another significant feature of the optimal allocation is that  the allocation (and therefore the performance) does not change for budgets larger than \$124. Even though $NVM_1$ has both faster read and write times than $NVM_2$, it is more favorable to keep 89\% of the disk pages in $NVM_2$, independent
of cost. The reason  is that failure costs play a significant role, despite the fact that the likelihood of a failure is very small. Disk blocks that are placed in $NVM_2$ under  the optimal placement have relatively low request frequency (averaging 3.4 requests over the course of the week-long trace), and
therefore, there is less advantage to having them in $NVM_1$. Even though failures are very unlikely, the difference in failure rate between $NVM_1$ and $NVM_2$ becomes significant in expectation when multiplied times the cost of recovering the  page from Disk in the event of a failure. By contrast, the disk pages allocated to $NVM_1$ have a much higher request rate (averaging 233 over the course of the trace) and the benefit of the faster response time of $NVM_1$ more than offsets the increased probability of having to retrieve  them from Disk in the case of a failure.

In a single server system, it may not make sense to average over the effect of events like failures
that happen once every few years even if they are significant in expectation. On the other hand, with a multi-node cache configuration consisting of a large number of servers, the likelihood of failures in a shorter period of time is much higher and it is sensible to average in their impact.

\begin{figure}
    \centering\small
        \includegraphics[width=8cm]{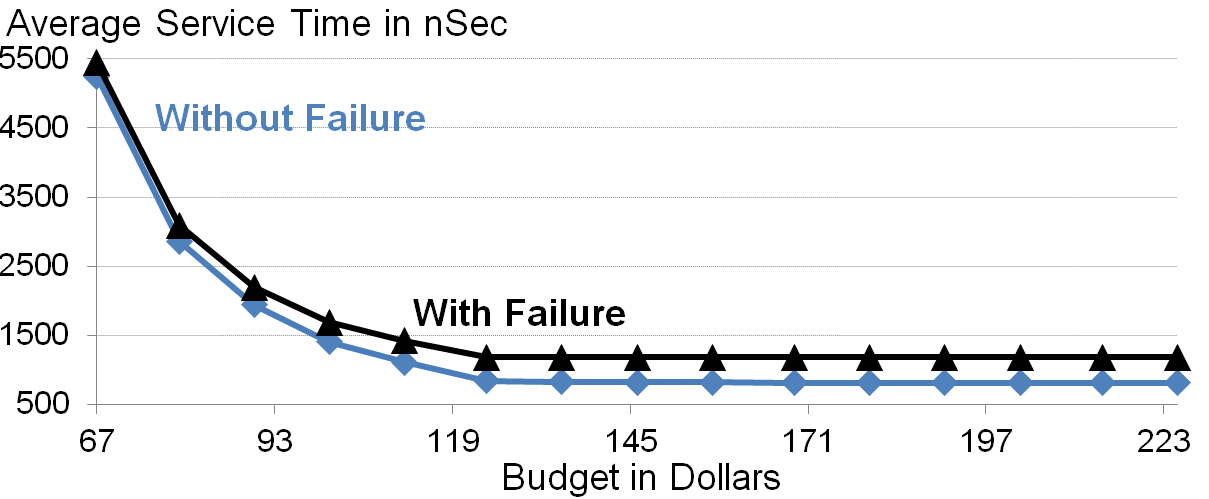}
\caption{The average cost to service requests under the optimal cache configuration for different budgets.}
\label{fig:MailTierFailNoFailComp}
\end{figure}

Figure \ref{fig:MailTierAllocNoFail} shows the same experiment run except that the cost of failures is not included in the service time. The service time for a disk page is just the expected time servicing read and write requests to the page. With the cost of failures removed, the optimal placement (with unlimited budget) does have all of the disk pages in $NVM_1$. Although the allocation changes as the budget is increased, it's not clear whether the additional expenditure has a significant impact on the expected service time. Figure \ref{fig:MailTierFailNoFailComp} shows the expected service time 
under the optimal configuration and placement for each budget. One line corresponds to the first set of allocations in which the cost of restoring after failures is taken into account. The second line corresponds to the scenario in which failures are not included. The difference between the two lines is the expected cost of failures. 
The difference is more pronounced (higher than 40\% difference) with higher budgets as more disk pages are placed in 
either $NVM_1$ or $NVM_2$. The graph also indicates that performance does not improve significantly in either scenario past a budget of \$124. Thus, even though it is optimal with no failures to store everything in $NVM_1$ if the budget allows, the improvement past \$124 is not significant.

\begin{figure}
    \centering\small
        \includegraphics[width=8cm]{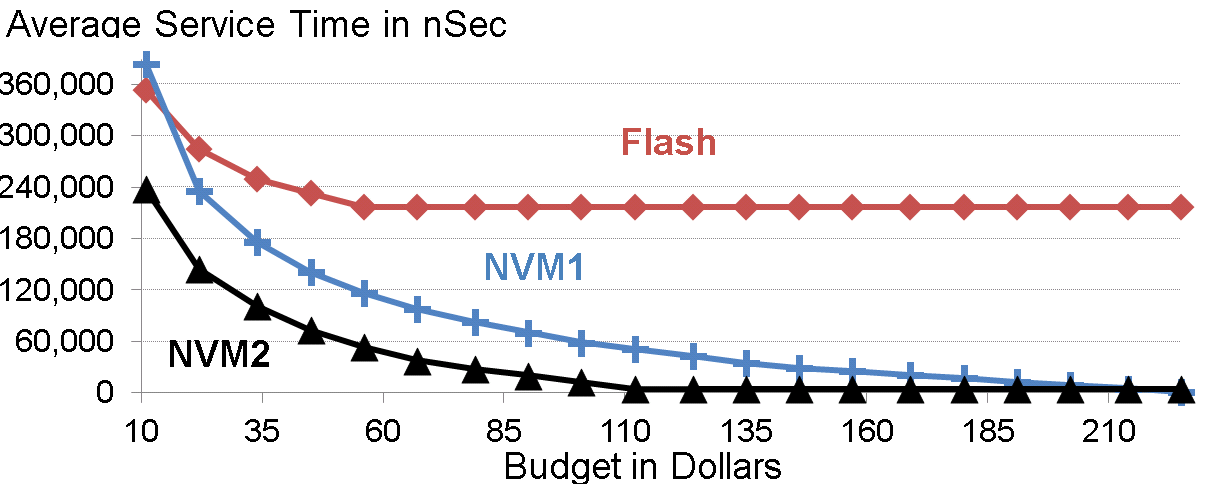}
\caption{Average service time with only one stash. 
}
\label{fig:MailTierOneType}
\end{figure}

\begin{figure}
    \centering\small
        \includegraphics[width=8cm]{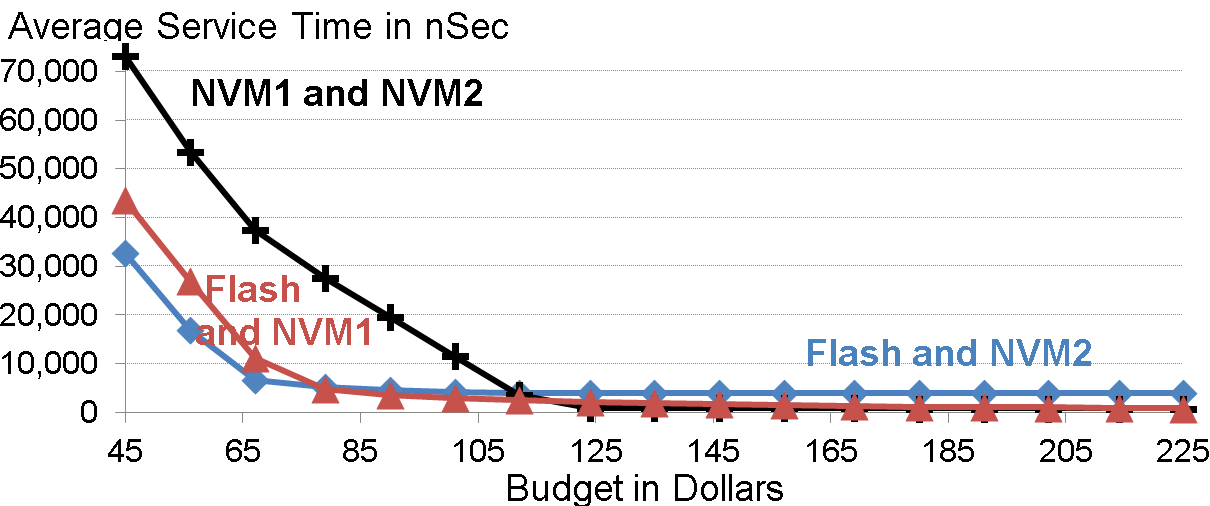}
\caption{The average cost to service requests when there are two stashes using a tiering policy. 
}
\label{fig:MailTierTwoTypes}
\end{figure}

It may be favorable to have a cache consisting of fewer stashes. 
If so, our approach can be
used to select a good set of memory types to include. 
In the next set of experiments we evaluate the impact on service time when  the set of memory types is restricted.
In Figure \ref{fig:MailTierOneType}, each curve represents an experiment in which there is exactly one stash.
A disk page can either be placed on that stash or excluded from the cache.
$NVM_2$ generally does better than $NVM_1$ for anything but the full \$225 budget because more disk pages
can fit in the cache. 
However, when the budget reaches the maximum \$225,
the average response time with $NVM_1$ is 807 nS as compared to 3900 with $NVM_2$.
In Figure \ref{fig:MailTierTwoTypes}, each curve represents an experiment in which there are exactly
two stashes. A policy of tiering is employed, so a disk page can either be included in one of the two stashes or
excluded from the cache. A combination of Flash and $NVM_1$ does better for a broader range of budgets,
but $NVM_2$ and $NVM_1$ does better at the higher end of the scale.

\begin{figure}
    \centering\small
       \includegraphics[width=8cm]{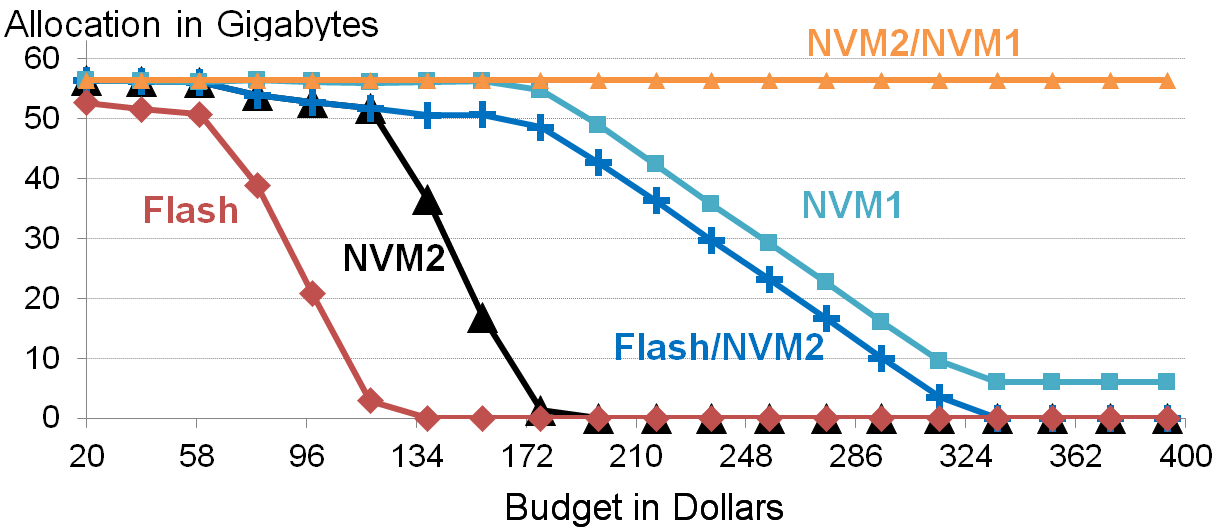}
\caption{The optimal partition of disk pages among stashes when all $8$ placements
are allowed.}
\label{fig:MailReplAlloc}
\end{figure}

Finally, we evaluate whether it is beneficial to allow some replication. Disk pages with low write rates 
will incur less overhead in having the additional copy on a less expensive but more reliable stash.
Figure \ref{fig:MailReplAlloc} is  the allocation of disk pages to stashes when all $8$ possible
placements on the three stashes is allowed. No disk pages were ever placed on all three caches.
The placement  option $\{\mbox{Flash}, NVM_1\}$ had a very small allocation but was removed from the graph because
it was too difficult to see. 

\begin{figure}
    \centering\small
        \includegraphics[width=8cm]{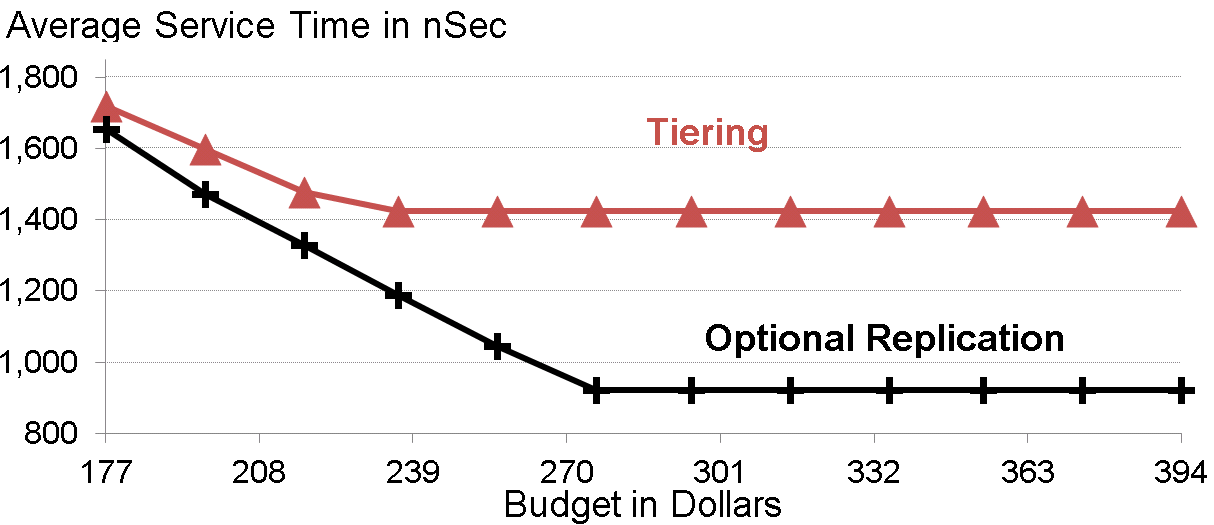}
\caption{Comparison of tiering and optional replication with two stashes: Flash and $NVM_1$. The average
service time for
forced replication is not shown because even at the very highest budget range, the average response time
was $212,459$ nS.}
\label{fig:MailTierReplComp}
\end{figure}

Finally, we compare tiering and replication with a cache that includes Flash and $NVM_1$.
We consider two variants of replication. Under {\em optional replication}, a key-value pair in $NVM_1$
may or may not also reside in Flash. Under {\em forced replication}, every key-value pair in $NVM_1$ 
must also have a copy in Flash. Naturally, the most flexible variant (optional replication) will be at
least as good as the other two policies (forced replication and tiering).

The graph in Figure \ref{fig:MailTierReplComp} shows that the  optimal placements under  tiering and optional replication do differ as there
are some disk pages that are written so infrequently that the cost  of maintaining the additional copy is offset by the 
expected cost of restoring a copy to $NVM_1$ in case of a failure.
The forced replication option is not even shown because it was very costly in comparison to the other
two policies. For the high budget range, the average response time for tiering and optional replication are
approximately  $1.4 \mu$s and $.9 \mu$s, respectively. The corresponding value for forced replication is $212 \mu$s.
Disk pages that are updated frequently incur a high cost for replication. 

\subsection{Key-Value Store Caches}
\label{sec:KVS}

Today's key-value store caches such as memcached use DRAM to store key-value pairs.
An instance loses its content in the presence of a power failure.
In this section, we consider a memcached instance that might be configured with 
five possible memory types for the cache: Disk, Flash, $NVM_2$, $NVM_1$, and DRAM.

Our evaluation employs 
traces from a cache augmented SQL system that processes social networking actions issued by the BG benchmark~\cite{sumita13}.
The mix of actions is 99\% read and 1\% write which  is typical of social networking sites such as Facebook~\cite{TAO13}. 
The  trace corresponds to approximately 40 minutes of requests in which there are 1.1 million requests to 564 thousand key-value pairs. 
The  total size of the key-value pairs requested is slightly less than 25 gigabytes. 
The cost of storing the entire database on the most expensive stash, DRAM, is just under \$200.

When a key-value pair is absent from the cache,  it must be recomputed
by issuing one or more queries to the SQL system after every read which references it.
The time for this computation is provided in the trace file. 
An update (write request) to a key-value pair is an update to the relational data used to compute that key-value pair. 
If the key-value pair is not stored on a stash, it does not need to be refilled (written to the cache). 

In all of the budget scenarios considered, the optimal placement never assigned a key-value pair
to Disk. 
This is because
the cost of reading the key-value pair from Disk was more
expensive than computing the key-value pair from the database. 
In other traces, there were some key-value pairs which were more expensive to compute than to retrieve from
Disk. 
However, Disk was not a viable option for many of these key-value pairs because the corresponding 
point was not on the convex hull of placement options. 
This is similar to the discussion of the
left graph of Figure~\ref{fig:OtherKeys} that illustrates a
scenario in which DRAM is not an option for a key-value pair.

\begin{figure}
    \centering\small
        \includegraphics[width=8cm]{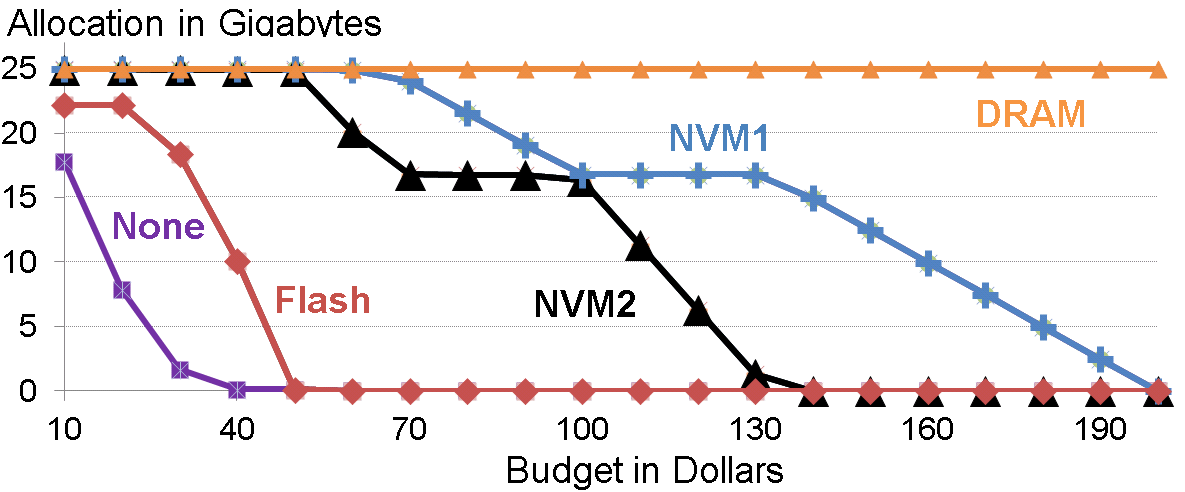}
\caption{The optimal partition of the key-value pairs among the stashes as the budget varies. The cost of stash failures is not included in the evaluation of optimality.}
\label{fig:BigBGTierAllocNoFail}
\end{figure}

Figure \ref{fig:BigBGTierAllocNoFail} shows the optimal size of each stash under a tiering policy  with all five memory types available as placement options. At each budget point, the vast majority of the key-value pairs were stored in three consecutive stashes which means that it was
generally more cost-effective to clear out key-value pairs from very slow stashes before investing in much faster space  for the
high-frequency items. 
Although not visible in the graph, under the optimal allocation, even for large budgets, there is approximately 8 MB of data that is not stored in the cache at all. These key-value pairs had one write request but no read requests over the course of the trace, so having those key-value pairs outside the cache reduced the average service time (although only slightly). The graph below shows the allocation in the scenario in which failure costs are not counted. When failures were counted, approximately 2/3 of the database was stored in $NVM_1$ instead of DRAM,
even at the high end of the budget range. These key-value pairs were read only once during  the entire trace in contrast with the key-value pairs stored in DRAM which were read on average about 4 times during the trace. Although it was slightly better to have the low frequency key-value pairs in $NVM_1$, the effect on the cost was almost negligible if  they were included in DRAM instead. 
This illustrates that  there can  be many substantially different placements that are all close to the optimal in their average service time.
These alternatives can be explored by limiting the set of placement options and comparing the average service time under the
more restrictive scenario to the  average service time in which all possibilities are allowed. 
The next set of experiments carry out this idea.

\begin{figure}
    \centering\small
        \includegraphics[width=8cm]{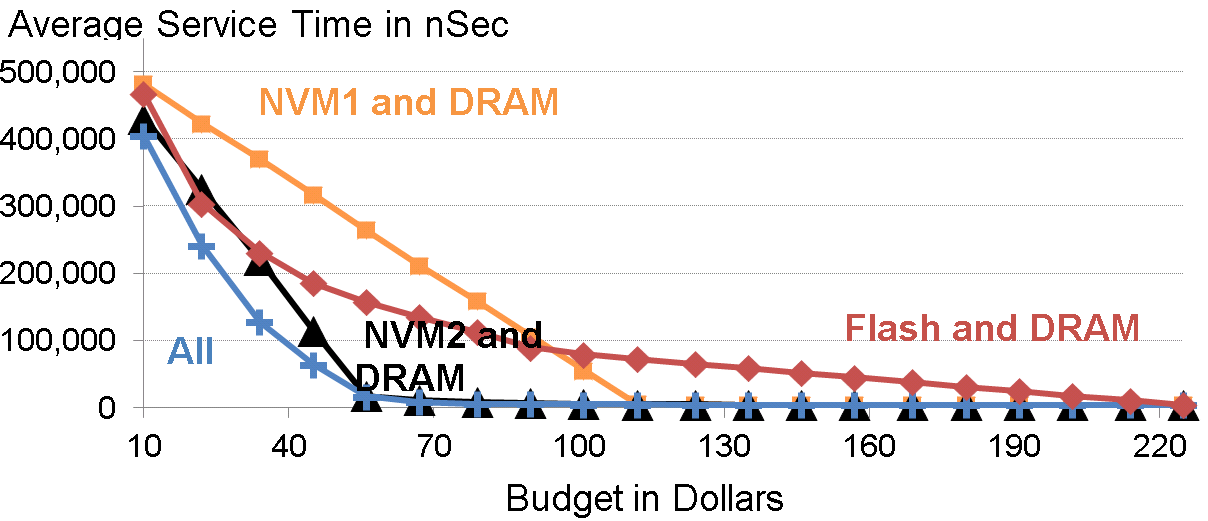}
   
        \caption{Comparison of cache configurations with two stashes with tiering.
        }
\label{fig:BigBGTierTwoTypesFull}
\end{figure}

We evaluate the cache performance when the cache consists of
DRAM in combination with different types of NVM. 
Figure \ref{fig:BigBGTierTwoTypesFull} shows the performance of the cache as a function of budget for the scenario where DRAM is combined with one other storage option. The combination that does well over the broadest range of budgets is $NVM_2$ and DRAM.
$NVM_1$ and DRAM do the best at the highest price range, but $NVM_2$ and DRAM is very close.
In Figure~\ref{fig:KVSTieringPerf1choice}, where the cache is limited to
one stash, $NVM_2$ provides the best service time with budgets lower than \$100.

\begin{figure}
    \centering\small
        \includegraphics[width=8cm]{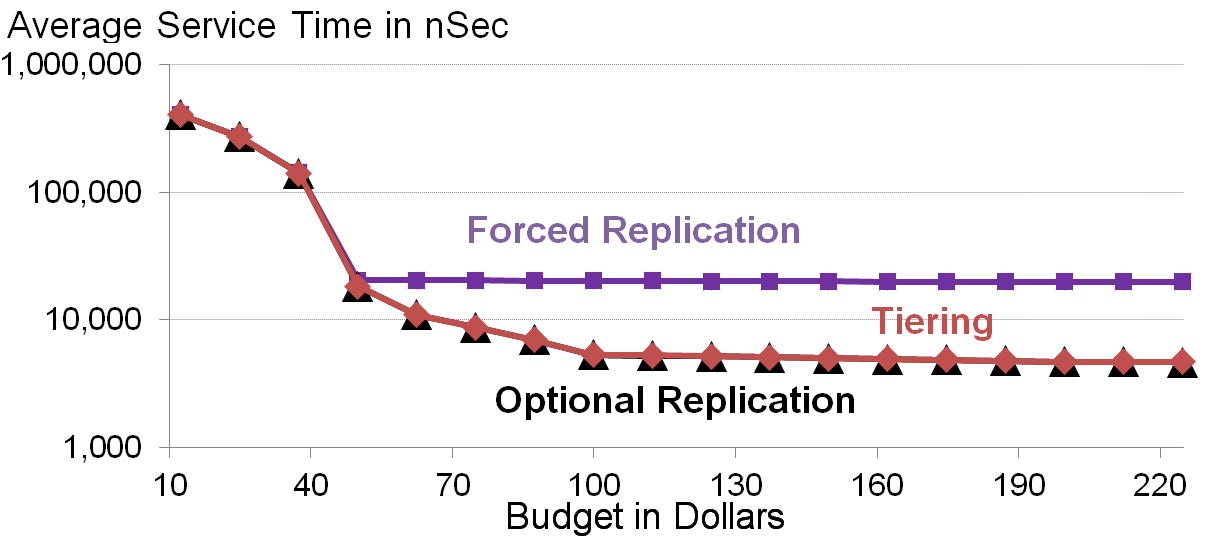}
\caption{A comparison of tiering and replication policies with a cache that includes $NVM_2$ and DRAM.
Under optional replication, a key-value pair can reside in $NVM_2$, DRAM or both. Under forced replication,
every key-value pair in DRAM is also in $NVM_2$. In tiering, a key-value pair can only reside in one stash. The lines 
for Optional Replication and Tiering are extremely close.}
\label{fig:BigBGTierReplComp}
\end{figure}

Finally, we compare tiering and replication with a cache that includes  $NVM_2$ and DRAM.
The graph in Figure \ref{fig:BigBGTierReplComp} shows the data for tiering, optional replication
and forced replication.
Forced replication is significantly worse than 
the other two as the cost of updating key-value pairs in both stashes is expensive.
The optimal placements under  tiering and optional replication do differ slightly as the set of
key-value pairs that are never updated are stored on both $NVM_2$ and DRAM under optimal replication.
However,
the difference in performance is so negligible that the two lines cannot be distinguished in the graph. 
This data is more evidence that under this request distribution, the impact of memory failures
is not significant and that tiering is a good choice for allocating key-value pairs to stashes.

\section{Related Work}\label{sec:related}

An overview of the different types of memory including NVM is provided in~\cite{adapt2014}.  This study motivates the development
of both offline and online algorithms for managing storage for database applications, but it does not present specific algorithms.

Several studies have investigated a multi-level cache hierarchy in the context of distributed file servers~\cite{muntz92,zhou2004}.  These studies observe that LRU may not work well for the intermediate caches and present alternative online algorithms.  The concept of {\em inclusive} and {\em exclusive} cache hierarchies is presented in~\cite{wong2002}.  
Inclusive provides for duplication of disk blocks (similar to replication of data items) while exclusive de-duplicates blocks across the caches (tiering of data items).  The approach in ~\cite{wong2002}
is to extend LRU with a demote operation to implement an exclusive cache.  None of these studies configure a cache by selecting the storage mediums that should participate as a stash in the multi-level hierarchy.  Novel features of our proposed approach include its optimality in serving as a measuring yardstick to evaluate alternative on-line algorithms and its consideration of failure rates.

A cache hierarchy consisting of PCM and NAND Flash is analyzed in~\cite{kimpcm2014}.  While the focus of this study is on PCM and its viability as a stash for use as a host-side cache, it presents an offline algorithm to tier 1 GB extents (consisting of 4 Kilobyte disk pages) across a hierarchy composed of PCM, Flash, and Disk.  They evaluate the performance of a  cache configuration in terms of I/Os per second (IOPS). Their method exhaustively searches all possible combinations of PCM, Flash, and Disk to find the one that maximizes 
 (IOPS)/\$.  They use a heuristic to place data items into a candidate cache configuration 
in order to evaluate its performance (IOPS).
We implemented their  data placement algorithm and found
that it did in fact find the optimal placement on the traces in our study. However, it is also possible to devise workload scenarios in which 
the data placement algorithm from~\cite{kimpcm2014} is
provably sub-optimal. An example is given in the appendix.
Our approach is a superset of theirs as it considers both tiering and replication with an arbitrary mix of storage medium and failure rates.  Moreover, our method simultaneously optimizes cache configuration and placement subject to a budget constraint.
Finally, our method is provably optimal and can be used as a measuring yardstick to evaluate heuristics.

\section{Future Work}\label{sec:future}
An important next step is to evaluate online replacement policies in conjunction with the
offline cache configuration method proposed here.  This requires an extension of our model
to consider the overhead of moving data items between the stashes.  A simple approach may
employ a static placement that is recomputed and updated periodically as the popularity
of different items vary over time~\cite{kimpcm2014}.
Alternatively, a cache replacement policy might be a
variant of online algorithms for two-level hierarchies such as CAMP~\cite{camp14}.

Another  important  direction  to  pursue  is  to  evaluate  how  robust  a  configuration  design  is  to changes  in  the  workload characteristics.   
The  use  of  past  statistics  is  bound  to  be  only an approximation of the workload in the future.  Therefore, it is important to understand
how well a particular configuration behaves as the set of data items grows over time or as the characteristics  of  the  access  pattern  change  and  when  it  is appropriate  to  alter  the  configuration.

\bibliographystyle{acm}
\bibliography{cit}  

\section*{Appendix}

This section gives an example showing the non-optimality of the placement algorithm used in \cite{kimpcm2014}.
We use the same numbers for the device latencies as given in \cite{kimpcm2014} and as written in the table below:

\vspace{.1in}
\begin{center}
\begin{tabular}{rrrr}
\hline
& PCM & Flash  & 15K Disk\\
\hline
$T_R$ = 4 Ki R. Lat. & 6.7 $\mu$s & 108.0 $\mu$s & 5000 $\mu$s\\
$T_W$ = 4 Ki W. Lat. & 128.3 $\mu$s & 37.1 $\mu$s & 5000 $\mu$s\\
\hline
\end{tabular}
\end{center}
\vspace{.2in}

The sample database is small with only two items. The whole example can be scaled by replicating each item $n$
times and dividing all the frequencies by $n$.
The space allocation is that there is enough PCM and Flash  to each hold one of the two items, so neither
will be stored on Disk.
The frequencies are chosen so that the read rate is slightly higher than the write rate. The numbers can be 
altered to have different read-to-write ratios  so that the example will still hold.
\begin{eqnarray*}
f_R(1) & = & 2.4 N\\
f_W(1) & = & 2N\\
f_R(2) & = & 3.3N\\
f_W(2) & = & N\\
\end{eqnarray*}
$N$ is a normalizing constant so that the four values above all sum to 1.
Since all the numbers for the rest of the example have a factor of $N$, we will drop the
factor of $N$.
The algorithm defines the following three values for each item:
\begin{eqnarray*}
Score_{PCM} &  = & f_R \cdot (T_R(Disk) - T_R(PCM))  \\
& + &  f_W \cdot (T_W(Disk) - T_W(PCM)) \\
Score_{Flash} &  = &  f_R \cdot (T_R(Disk) - T_R(Flash)) \\
& + &f_W \cdot  (T_W(Disk) - T_W(Flash)) \\
Score &  = & \max\{ Score_{PCM}, Score_{Flash} \}
\end{eqnarray*}
According to these definitions, both items prefer PCM:
\begin{eqnarray*}
Score_{PCM}(1) &  = & 2.4 (5000 - 6.7) + 2 (5000 - 128.3) \\
& = & 21721\\
Score_{Flash}(1) &  = & 2.4 (5000 - 108) + 2 (5000 - 37.1)\\
& = & 21667\\
Score(1) &  = & 21721
\end{eqnarray*}
\begin{eqnarray*}
Score_{PCM}(2) &  = & 3.3 (5000 - 6.7) + (5000 - 128.3)\\
& = & 21350\\
Score_{Flash}(2) &  = & 3.3 (5000 - 108) + (5000 - 37.1)\\
& = & 21107\\
Score(2) &  = & 21350
\end{eqnarray*}
The algorithm of~\cite{kimpcm2014} orders the items according to their score.
The item with highest score (item $1$) goes first and is placed in its first choice location (PCM).
Then item $2$ is placed in Flash.
The overall expected time for an I/O is:
\begin{eqnarray*}
& & f_R(1) \cdot ReadLat_{PCM} + f_W(1) \cdot WriteLat_{PCM} \\
& + & f_R(2) \cdot ReadLat_{Flash} + f_W(2) \cdot WriteLat_{Flash}\\
& = & 2.4 \cdot 6.7   + 2 * 128.3 + 3.3 * 108  + 37.1\\
& = & 666.18
\end{eqnarray*}
According to the alternative placement, with item $1$ in Flash and item $2$ in PCM, 
the expected time for an I/O is:
\begin{eqnarray*}
& & f_R(2) \cdot ReadLat_{PCM} + f_W(2) \cdot WriteLat_{PCM} \\
& + & f_R(1) \cdot ReadLat_{Flash} + f_W(1) \cdot WriteLat_{Flash}\\
& = & 3.3 \cdot 6.7 + 128.3 + 2.4 * 108  + 2 \cdot 37.1\\
&= & 484.81
\end{eqnarray*}
The placement not chosen by the algorithm has a significantly lower 
expected time per I/O.

\end{document}